\documentclass[12pt]{article}
\usepackage{graphicx}
\usepackage{latexsym}
\advance\oddsidemargin by -\textwidth
\advance\oddsidemargin by -1in
\evensidemargin=\oddsidemargin
\def\fsl#1{\setbox0=\hbox{$#1$}           % set a box for #1 
   \dimen0=\wd0                                 % and get its size
   \setbox1=\hbox{/} \dimen1=\wd1               % get size of /
   \ifdim\dimen0>\dimen1                        % #1 is bigger
      \rlap{\hbox to \dimen0{\hfil/\hfil}}      % so center / in box
      #1                                        % and print #1
   \else                                        % / is bigger
      \rlap{\hbox to \dimen1{\hfil$#1$\hfil}}   % so center #1
      /                                         % and print /
   \fi}                                         %
\makeatletter
\def\@maketitle{\newpage
 \null
 {\normalsize \tt \begin{flushright} 
  \begin{tabular}[t]{l} \@date 
  \end{tabular}
 \end{flushright}}
 \begin{center}
 \vskip 2em
 {\LARGE \@title \par} \vskip 2.5em {\large \lineskip .5em
 \begin{tabular}[t]{c}\@author 
 \end{tabular}\par} 
 \end{center}
} 

\newcommand{\vereq}[2]{\lower3pt\vbox{\baselineskip1.5pt \lineskip1.5pt
\ialign{$\m@th#1\hfill##\hfil$\crcr#2\crcr\sim\crcr}}}

\newcommand\eqsecnum{
\@newctr{equation}[section]
\renewcommand\theequation{\arabic{section}.\arabic{equation}}%
}

\newbox\tempboxa
\newdimen\captionboxsubcount 
\def\capsize#1{\captionboxsubcount=#1pt}
\newdimen\captionboxsub
\captionboxsub=\hsize \advance\captionboxsub by -\captionboxsubcount
\advance\captionboxsub by -\captionboxsubcount
\long\def\@makecaption#1#2{
 \setbox\@tempboxa\hbox{\footnotesize #1: #2}
 \ifdim \wd\@tempboxa >\captionboxsub 
\rightskip=\captionboxsubcount \leftskip=\captionboxsubcount 
  \footnotesize #1: #2 
\else \hbox to\hsize{\hfil\box\@tempboxa\hfil} 
 \fi}
\makeatother
\eqsecnum
\capsize{30}
\title{{\Large\bf 
   $\pi^+ - \pi^0$ mass difference\\ from the Bethe-Salpeter equation
\vspace{5mm}}}
\author{
{\large
    Masayasu {\sc Harada}\thanks{
      {\tt harada@eken.phys.nagoya-u.ac.jp}}, \ \ 
    Masafumi {\sc Kurachi}\thanks{
      {\tt kurachi@eken.phys.nagoya-u.ac.jp}} \ \  and \ \ 
    Koichi {\sc Yamawaki}\thanks{
      {\tt yamawaki@eken.phys.nagoya-u.ac.jp}}
  }\\[3mm]
  {\it Department of Physics, Nagoya University}\\
  {\it Nagoya 464-8602, Japan}\\[1cm]
}
\date{
  DPNU-04-01 \\   \today\vspace{1cm}\\
}
\begin{document}
\maketitle

%%%%%%%%%%%%%%%%%%%%%%%%%%%%%%%%%%%%%%%%%%%%%%%%%%%%%%%%%%%%%%%%%%%%
%%%%%%%%%%%%%%%%%%%%%%%%%%%%%%%%%%%%%%%%%%%%%%%%%%%%%%%%%%%%%%%%%%%%
\vspace{1cm}
\begin{abstract}
In the framework of the Schwinger-Dyson equation and 
the Bethe-Salpeter equation in the 
improved ladder approximation, 
we calculate the $\pi^+ - \pi^0$ mass difference
on the same footing as
the pion decay constant and the 
QCD $S$ parameter (or $L_{10}$) through
the difference between 
the vector current correlator $\Pi_{VV}$ 
and the axial-vector current correlator $\Pi_{AA}$.
We find that all the results can be fit to the 
experimental values  for rather large 
$\Lambda_{\rm QCD} \sim 700 {\rm MeV}$
which reflects the ``scale ambiguity''. 
By fitting to the calculated data 
using the pole saturated form of $\Pi_{VV} - \Pi_{AA}$,
we also derive masses and decay constants of 
$\rho$ meson and $a_1$ meson, which we found are 
consistent with the experiments rather 
insensitively to the ``scale ambiguity''.
\end{abstract}

%%%%%%%%%%%%%%%%%%%%%%%%%%%%%%%%%%%%%%%%%%%%%%%%%%%%%%%%%%%%%%%%%%%%
%%%%%%%%%%%%%%%%%%%%%%%%%%%%%%%%%%%%%%%%%%%%%%%%%%%%%%%%%%%%%%%%%%%%
\newpage
\section{Introduction}
\label{sec:Intro}

The $\pi^+$ - $\pi^0$ mass difference $\Delta m_\pi^2
\equiv  m_{\pi^+}^2 -m_{\pi^0}^2$ is an interesting quantity to
measure an explicit breaking of the chiral symmetry
by the gauge coupling ($U(1)_{\rm em}$) in the spontaneously 
broken phase of the
chiral symmetry. 
It in fact has been a prototype of the
mass calculation of pseudo Nambu-Goldstone (NG) bosons 
in strong coupling gauge theories 
such as those 
in the technicolor theories~\cite{techni}
and more recently in the little Higgs 
models~\cite{Arkani-Hamed:2001nc}.
The sign as well as the absolute value of $\Delta m_\pi^2$
is an important issue (``vacuum alignment problem''), 
since the negative sign would imply
that the vacuum would align so as to break spontaneously the gauge symmetry 
($U(1)_{\rm em}$ in 
the case of pion) which explicitly breaks the spontaneously broken 
symmetry. Such a situation never happens in the
real-life QCD but may do in other theories.
Hence this type of calculation plays a central role of the model
buildings. 

The first successful 
calculation of $\Delta m_\pi^2$~\cite{Das:it}  was done by the
current algebra 
in conjunction with the Weinberg
spectral function sum rules~\cite{Weinberg:1967kj}
saturated by the  $\pi$, $\rho$ and $a_1$ meson poles.
Somewhat more elaborate calculation on this line
was done combined with the QCD 
information~\cite{Bardeen}. Recently, an 
effective field theory calculation without $a_1$ meson has been 
successfully done~\cite{Harada:2003xa}, 
based on the Hidden Local Symmetry (HLS) 
model~\cite{BKUYY,BKY:NPB,BKY:PTP} at loop 
level~\cite{HY:92,Tanabashi:93}
and the Wilsonian matching~\cite{HY:WM}  
of the HLS model with the QCD (for reviews
of the HLS approach, see \cite{BKY,Harada:2003jx}).
Ref. \cite{Harada:2003xa} demonstrated that the HLS model is a little
Higgs 
model with two sites and two links (open moose) whose 
locality of theory space forbids quadratic divergence
in $\Delta m_\pi^2$ even without $a_1$ meson.

On the other hand, direct QCD calculation of the
$\Delta m_\pi^2$ has never been done except for
the lattice simulation~\cite{Duncan:1996xy}.
Although
the lattice QCD is a powerful method,
it is still important 
to establish alternative QCD-based
methods which are more intuitive and
less dependent on the power of computers.
Such methods will be crucial
to the subjects which are difficult to be studied in the framework of 
the lattice simulation. Most relevant subjects are
the dynamical origin of the 
electroweak symmetry breaking,
QCD at finite density, etc.

In this paper, we calculate in the chiral
symmetric limit the $\Delta m_\pi^2$  simultaneously with
the the pion decay constant $f_\pi$ and the QCD $S$ parameter~\cite{PeskinTakeuchi} 
(or the Gasser-Leutwyler parameter $L_{10}$~\cite{GL})
within the framework of the
the Schwinger-Dyson (SD) equation and 
the Bethe-Salpeter (BS) equation 
in the the improved ladder approximation.
We note  that 
all these quantities are given by ``generalized'' Weinberg spectral 
function sum rules for the difference between 
the vector current correlator $\Pi_{VV}$ 
and the axial-vector current correlator $\Pi_{AA}$: 
$f_\pi^2$ is given by the Weinberg first sum rule, the $S$ parameter
by the Das-Mathur-Okubo (DMO) sum rule~\cite{DMO} or the ``zeroth
Weinberg 
sum rule'', and $\Delta m_\pi^2$ by the Das-Guralnik-Mathur-Low-Young
(DGMLY) 
sum rule~\cite{Das:it} or the ``third Weinberg sum rule''. We then
rewrite them in terms of the current correlators instead of the
spectral  
functions and directly calculate 
$\Delta m_\pi^2 $ 
on the same footing as $f_\pi$ and the  
QCD $S$ parameter through
$\Pi_{VV} - \Pi_{AA}$ in the {\it space-like momentum region}.
Such correlators are written in terms of the BS amplitudes
which are given by solving 
the {\it inhomogeneous} BS (IBS) equation and the SD equation
with the improved ladder approximation in the Landau gauge
both in the {\it space-like region}.
In contrast to our case, solving the bound state spectra by the
homogenous  
BS equation in the time-like region is usually difficult,
since it would need
analytic continuation of the running coupling 
which is not an analytic function.~\footnote{
 See an exception in Ref.~\cite{Harada:2003dc}.
}

This kind of method has been extensively used 
to investigate masses and decay constants of 
low lying
mesons~\cite{Aoki:1990eq,Aoki:1990yp,Roberts:2000aa}
in good agreement with the experiments. On the other hand,
the same method failed to reproduce 
the QCD $S$ parameter consistently with 
the experiment~\cite{Harada:1994ni}. 

However, we find that the QCD $S$ parameter is rather sensitive to 
the infrared (IR) cutoff parameter of the QCD running coupling.
 We actually reproduce the QCD $S$ parameter 
consistently 
with the experiment by setting the IR cutoff parameter in the region 
never investigated in the previous works. 
Our parameter choice corresponds to taking
the QCD scale parameter
as $\Lambda_{\rm QCD} \simeq 724 \, {\rm MeV}$, while it
 was chosen as
$\Lambda_{\rm QCD} \simeq 500 \,{\rm MeV}$ in the previous results of
the decay constants and the masses~\cite{Aoki:1990eq,Aoki:1990yp} and
of the 
QCD $S$  parameter~\cite{Harada:1994ni}, 
both values being considerably higher than 
the conventional $\Lambda_{\rm QCD}^{(3)} = 300$-$450$\,MeV obtained
in  
the $\overline{\rm MS}$ scheme~\cite{Buras}.
This actually corresponds
to exploiting the ``scale ambiguity''~\cite{BLM} 
of $\Lambda_{\rm QCD}$ in
the SD and BS equations recently 
emphasized by Hashimoto and Tanabashi~\cite{Hashimoto:2002px}.

We then show that $\pi^+ - \pi^0$ mass difference 
$\Delta m_\pi^2$ is 
calculated in rough agreement with the experimental value for the same
parameter region as that reproducing the QCD $S$ parameter and $f_\pi$.
Actually, although infrared dynamics is quite important 
for determination of $f_\pi$ and $S$, it is  
not so important for determination of $\Delta m_\pi^2$.
In other words, $f_\pi$ and $S$ are sensitive to 
the infrared structure of the running coupling, while 
$\Delta m_\pi^2$ is not.

We also derive the masses and decay constants 
of $\rho$ meson and $a_1$ meson by fitting them to
the calculated 
$\Pi_{VV} - \Pi_{AA}$ within our method using the 
pole saturated form of $\Pi_{VV} - \Pi_{AA}$.
These quantities are also compared with experiments, which  
is another check of the validity of the calculation in the present
analysis. 

The calculation presented in this paper turns out to be the 
first example which derives $\Delta m_\pi^2$ directly from QCD 
without depending on the lattice simulation.
Once we have checked the reliability of such a method in the 
ordinary QCD (with three light flavors), we may  
apply it to the mass of pseudo NG bosons and 
vacuum alignment problem in
other strong coupling gauge 
theories such as the large $N_f$ QCD, walking technicolor, little
Higgs, 
hot/dense QCD, etc..

This paper is organized as follows.
In section~\ref{sec:Spectral} we introduce spectral function 
sum rules: The DMO sum rule or the zeroth Weinberg sum rule for $S$
parameter, the 
first Weinberg sum rule for $f_\pi^2$, the second Weinberg sum rule,
and  
the DGMLY sum rule or the third Weinberg sum rule 
for $\Delta m_\pi^2$. We then rewrite them in terms of 
the current correlators $\Pi_{VV} - \Pi_{AA}$.
In section~\ref{sec:correlator-BS} we show how the current
correlators are obtained from the BS amplitude which is
calculated from the IBS equation in the space-like region.
In section~\ref{sec:IBS} we introduce the IBS equation 
and show how to solve it numerically.
Section~\ref{sec:results} is the main part of this paper:
We first show the result of the IBS equation for QCD, 
and discuss the dependence of $f_\pi$, $S$ and $\Delta m_\pi^2$ 
on the infrared structure of the running coupling.
Then we compare them with the experimental values.
We also show results for masses and decay constants of 
$\rho$ and $a_1$ mesons.
In section~\ref{sec:summary} we give summary and 
discussions  on applications 
to the large $N_f$ QCD and the electroweak 
symmetry breaking.

%%%%%%%%%%%%%%%%%%%%%%%%%%%%%%%%%%%%%%%%%%%%%%%%%%%%%%%%%%%%%%%%%%%%
%%%%%%%%%%%%%%%%%%%%%%%%%%%%%%%%%%%%%%%%%%%%%%%%%%%%%%%%%%%%%%%%%%%%
\section{Spectral function sum rules}
\label{sec:Spectral}

In this section we introduce the spectral function sum rules, 
which express the
pion decay constant $f_\pi$, the QCD $S$ parameter and 
$\Delta m_\pi^2$ in terms of $\Pi_{VV} - \Pi_{AA}$
in the space-like momentum region. Here we consider only the chiral
symmetric
limit of massless three flavors ($N_f=3$), 
the corrections of the finite quark mass being
expected to be small for the quantities we consider. 

Let us begin with
introducing the vector and axial-vector currents as 
\begin{equation}
  V_\mu^a(x) = \bar\psi(x) T^a \gamma_\mu \psi(x)\, ,
\end{equation}
\begin{equation}
  A_\mu^a(x) = \bar\psi(x) T^a \gamma_\mu \gamma_5
  \psi(x)\, ,
\end{equation}
where 
$T^a$ is the generator of $SU(N_f)$ normalized as 
$\mbox{tr}(T^a T^b) = \frac{1}{2}
\delta^{ab}$ and we consider $a,b=1,2,3$.
In the chiral symmetric limit the current correlator $\Pi_{JJ}$ is
given as 
\begin{eqnarray}
  \delta^{ab} \left( \frac{q_\mu q_\nu}{q^2} - g_{\mu \nu} \right) 
\Pi_{JJ}(q^2) &=& i 
  \int d^4x\ e^{i q x}\ \langle 0 \vert T J_\mu^a(x) J_\nu^b(0) 
  \vert 0 \rangle\, , \\ 
& & \hspace{2cm} 
    ( J_\mu^a(x) = V_\mu^a(x), A_\mu^a(x) )\, . \nonumber
\end{eqnarray}
The Umezawa-Kamefuchi-K\"allen-Lehmann spectral representation 
for the current correlators are expressed as
\begin{eqnarray}
  & & i  
  \int d^4x\ e^{i q x}\ \langle 0 \vert T J_\mu^a(x) J_\nu^b(0) 
  \vert 0 \rangle \nonumber \\
  & & \hspace{7mm} =   
  - \delta^{ab} \int_{0}^{\infty} ds 
  \frac{1}{s-q^2-i\epsilon} 
  \left\{ \left( g_{\mu \nu} - \frac{q_\mu q_\nu}{s} \right)
  \rho_{J}^{(1)}(s) - q_\mu q_\nu \rho_{J}^{(0)}(s) 
  \right\}\, .
\end{eqnarray}
Here, we decomposed the spectral function into the spin-one 
part $\rho_{J}^{(1)}(s)$ and spin-zero part $\rho_{J}^{(0)}(s)$.
Since the axial-vector current $A_\mu^a$ couples 
to the massless NG
boson, the pion, while no massless particles couple to the 
vector current $V_\mu^a$, we have 
\begin{eqnarray}
  \rho_{V}^{(0)}(s) &=& 0,\\
  \rho_{A}^{(0)}(s) &=& f_\pi^2 \delta (s),
\end{eqnarray}
where $f_\pi$ is the decay constant of the NG boson $\pi$ defined by 
\begin{equation}
  \langle 0 \vert A_\mu^a (0) \vert \pi^b(q) \rangle 
  = i q_\mu f_\pi \delta^{ab}\, .
\end{equation}

Using the spectral functions, we can write down the 
following ``generalized'' Weinberg sum rules:
\begin{eqnarray}
  \int_{0}^{\infty} \frac{ds}{s^2}\ 
  \left[ \rho_{V}^{(1)}(s) - \rho_{A}^{(1)}(s) \right] 
  &=& \frac{S}{4\pi}\, ,
  \label{eq:zeroth}\\
  \int_{0}^{\infty} \frac{ds}{s}\ 
  \left[ \rho_{V}^{(1)}(s) - \rho_{A}^{(1)}(s) \right] 
  &=& f_\pi^2\, ,
  \label{eq:first}\\
  \int_{0}^{\infty} ds\ 
  \left[ \rho_{V}^{(1)}(s) - \rho_{A}^{(1)}(s) \right] 
  &=& 0\, ,
  \label{eq:second}\\
  - \frac{3 \alpha_{em}}{4\pi f_\pi^2} 
  \int_{0}^{\infty} ds \log{s}\ 
  \left[ \rho_{V}^{(1)}(s) - \rho_{A}^{(1)}(s) \right] 
  &=& \Delta m_\pi^2\, ,
  \label{eq:third}
\end{eqnarray}
where, $S$ is the QCD $S$ parameter~\cite{PeskinTakeuchi}, which 
in the case of $N_f = 3$ is related to the
Gasser-Leutwyler parameter $L_{10}$~\cite{GL} as 
$S = -16\pi L_{10}$, 
$\alpha_{em} (= \frac{e^2}{4 \pi} = \frac{1}{137})$ 
is the coupling constant of the electromagnetic interaction, 
and $\Delta m_\pi^2$ is the $\pi^+-\pi^0$ mass difference 
defined by  
$\Delta m_\pi^2 \equiv m_{\pi^+}^2 - m_{\pi^0}^2$.
Equations~(\ref{eq:first}) and (\ref{eq:second}) are 
the first and second Weinberg sum rules~\cite{Weinberg:1967kj}, 
respectively, and Eqs.~(\ref{eq:zeroth}) is the DMO sum
rule~\cite{DMO} or    
often called the ``zeroth Weinberg
sum rule''. 
Equation~(\ref{eq:third}) was derived by 
Das et al.~\cite{Das:it} 
 and may be called the
DGMLY sum rule or the ``third Weinberg sum rule''.

Of course these relations in Eqs.~(\ref{eq:zeroth}), 
(\ref{eq:first}), (\ref{eq:second}), and (\ref{eq:third}) 
make sense only if the integrals converge.
For example, convergence of the second sum rule (\ref{eq:second}) 
requires that the difference between $\Pi_{VV}$ and $\Pi_{AA}$ 
must satisfy 
\begin{equation}
  Q^2 \left[ \Pi_{VV}(Q^2) - \Pi_{AA}(Q^2) \right]\ \ 
  \stackrel{( Q^2 \rightarrow \infty )}{\longrightarrow}\ \ 0\, ,
\label{eq:convergence}
\end{equation}
where $Q^2$ is related to the space-like momentum as 
$Q^2 = - q^2 ( >0 )$ .

The asymptotic behavior of the current correlators 
can be calculated
by the operator product expansion (OPE) technique.
In the chiral limit, the form of $\Pi_{VV}(Q^2) - \Pi_{AA}(Q^2)$ 
in the ultraviolet region is estimated 
as follows~\cite{Bernard,Shifman:bx}:
\begin{equation}
  \Pi_{VV}(Q^2) - \Pi_{AA}(Q^2) \ \ 
  \stackrel{( Q^2 \rightarrow \infty )}{\longrightarrow}\ \ 
  \frac{4\pi(N_c^2 - 1)}{N_c^2}\ 
  \frac{\alpha_s \langle \bar{q} q \rangle^2}{Q^4} 
\, ,
\label{eq:OPE}
\end{equation}
up to logarithm.
This means that $\Pi_{VV}(Q^2) - \Pi_{AA}(Q^2)$ indeed
satisfies 
the condition (\ref{eq:convergence}) and the relation in 
Eq.~(\ref{eq:second}) makes sense.
When the integral in the second sum rule converges, 
those in the first and zeroth sum rules are also convergent.

Now we rewrite the zeroth and the first sum rules  
in terms of the current correlators as 
\begin{eqnarray}
  S &=& \left.
  - 4 \pi \frac{d}{d Q^2} 
  \left[ \Pi_{VV}(Q^2) - \Pi_{AA}(Q^2)
  \right] \right\vert_{Q^2 = 0}\, ,
\label{eq:S_parameter}\\
  f_\pi^2 &=& \Pi_{VV}(0) - \Pi_{AA}(0)\, ,
\label{eq:fpi}
\end{eqnarray}
and also the third sum rule:~\cite{Yama} 
\begin{equation}
  \Delta m_\pi^2 
  = \frac{3 \alpha_{em}}{4 \pi f_\pi^2} 
    \int_{0}^{\infty} d Q^2 \left[ \Pi_{VV}(Q^2) -
  \Pi_{AA}(Q^2) \right]\, .
\label{eq:mass_diff}
\end{equation}
{}From Eq.(\ref{eq:mass_diff}), we can see that the asymptotic
behavior 
(\ref{eq:OPE}) also guarantees convergence of the third sum rule.

In Ref.~\cite{Harada:1994ni}
the pion decay constant $f_\pi$ and the QCD $S$ parameter
were calculated by the use of the above formulas.
However, 
$\Delta m_\pi^2$ was not calculated so far.
So the calculation presented in this paper turns out to be the 
first example which derives $\Delta m_\pi^2$ directly from QCD 
without depending on the lattice simulation.
Moreover, there is another interest regarding $\Delta m_\pi^2$ 
related to the structure of the QCD vacuum, as we stressed in the
Introduction.
When we switch off the electromagnetic interaction and set masses 
of quarks to be zero, pions are identified with the exact 
NG bosons associated with the spontaneous
breaking of 
$SU(2)_L \otimes SU(2)_R$ chiral symmetry 
down to $SU(2)_V$ symmetry.
The existence of the electromagnetic interaction explicitly 
breaks the $SU(2)_L \otimes SU(2)_R$ chiral symmetry, which makes
$\pi^+$ and $\pi^-$ be the pseudo NG bosons.
On the other hand, $\pi^0$ remains massless  
since the photon does not interact with $\pi^0$.
The interesting point here is whether 
$\Delta m_\pi^2 \equiv m_{\pi^+}^2 - m_{\pi^0}^2$ 
($= m_{\pi^+}^2$ in the chiral limit) becomes 
positive or negative, which is called ``vacuum alignment problem''.
Negative $\Delta m_\pi^2$ means that fluctuation of $\pi^+$ field 
around $\langle\pi^+\rangle = 0$ is unstable and 
the vacuum with 
$\langle\pi^+\rangle = 0$ is not a true vacuum.
If this is the case, $\pi^+$ has non-zero vacuum expectation value 
and $U(1)_{\rm em}$ symmetry is broken.
We know that, in the real world, $\Delta m_\pi^2$ is positive and 
the vacuum with 
$\langle\pi^+\rangle = 0$ is the true vacuum.
$U(1)_{\rm em}$ symmetry is not broken in the real world.
This is quite nontrivial fact resulting from 
the nonperturbative dynamics of the strong interaction.
So it is interesting to investigate whether we can reproduce 
positive $\Delta m_\pi^2$
which is realized in the real-life QCD.

%%%%%%%%%%%%%%%%%%%%%%%%%%%%%%%%%%%%%%%%%%%%%%%%%%%%%%%%%%%%%%%%%%%%
\section{Current correlators from BS amplitudes}
\label{sec:correlator-BS}

In the previous section we have written down the QCD $S$ parameter, 
$f_\pi$, and $\Delta m_\pi^2$ in terms of the current correlators.
Then, in this section, we show how the current correlators 
are obtained from the BS amplitude which will be
calculated from the IBS equation.

To derive properties of hadrons as boundstates, it is straightforward
to 
perform calculations directly in the time-like momentum region.
However, it is difficult to solve the BS equation and the SD equation 
in the time-like region 
since we have to carry out the analytic continuation of the running
coupling from the space-like region to the time-like region.
In the case of QCD, 
the one-loop running coupling is not an analytic function,
so that 
we have to approximate it by some analytic function 
in order to perform the calculations in the time-like region.
However, here we need the BS amplitude only
for the space-like region in order to calculate the current
correlators 
for $\Delta m_\pi^2$, $f_\pi$ and the QCD $S$ parameter.

The BS amplitude $\chi^{(J)}$ ($J=V$, $A$)
is defined in terms of the three-point vertex function 
as follows:
\begin{equation}
  \delta_i^j \left( \frac{\lambda^a}{2} \right)_{f}^{f'} 
  \int \frac{d^4 p}{(2 \pi)^4} e^{- i p r} \chi^{(J)}_{\alpha
  \beta}(p;q,\epsilon) = 
  \epsilon^\mu \int d^4 x e^{i q x} 
  \langle 0 \vert T\ \psi_{\alpha i f}(r/2)\ 
  \bar\psi_\beta^{j f'}(-r/2)\ J_\mu^a(x)\ 
  \vert 0 \rangle,
\label{eq:three-point}
\end{equation}
where $q^\mu$ is the total momentum of the fermion and
anti-fermion and $p^\mu$ is the relative one.
$\epsilon^\mu$ is the polarization vector defined by 
$\epsilon \cdot q = 0$, $\epsilon \cdot \epsilon = -1$, 
and $(f, f'), (i, j), (\alpha, \beta)$ are flavor, color and 
spinor indices, respectively.
Closing the fermion legs of 
the above three-point vertex function 
and taking the limit $r \rightarrow 0$, 
we can express the current correlator in terms of the 
BS amplitude as follows:
\begin{equation}
  \Pi_{JJ}(q^2) = \frac{1}{3} \sum_{\epsilon} \int \frac{d^4 p}{i (2
  \pi)^4} \frac{N_c}{2} \mbox{tr} \left[ 
  \left(\epsilon \cdot G^{(J)}\right) 
  \chi^{(J)}(p;q,\epsilon) \right]
\label{eq:Pi_JJ}
\end{equation}
where 
\begin{equation}
  G_\mu^{(V)} = \gamma_\mu,\ \ \ \ G_\mu^{(A)} = \gamma_\mu \gamma_5,
\end{equation}
and $N_c = 3$ is the number of colors.
In the above expression we averaged over the polarizations 
so that $\Pi_{JJ}(q^2)$ does not depend on the polarization.

We expand the BS amplitude  
$\chi_{\alpha \beta}^{(J)}(p;q,\epsilon)$  
in terms of the bispinor bases 
$\Gamma^{(J)}_i$ and the invariant amplitudes $\chi^{(J)}_i$ 
as 
\begin{equation}
  \left[ \chi^{(J)}(p;q,\epsilon) \right]_{\alpha \beta} 
  =\ \sum_{i=1}^{8} \left[ \Gamma_i^{(J)}(p;\hat{q},\epsilon)
  \right]_{\alpha \beta} \chi^{(J)}_{i} (p;q) ,
\label{eq:chi-expanded}
\end{equation}
where $\hat{q}_\mu = q_\mu/\sqrt{Q^2}$.
The bispinor bases can be chosen in a way that they have the same
properties of 
spin, parity 
and charge conjugation as the corresponding
current $J_\mu^a(x)$ has.
We adopt the following bispinor bases for the vector vertex:
\begin{eqnarray}
& &  \Gamma^{(V)}_1 = \fsl{\epsilon} ,\ \ 
     \Gamma^{(V)}_2 = \frac{1}{2} [\fsl{\epsilon},\fsl{p}] 
                           (p \cdot \hat q) ,\ \ 
     \Gamma^{(V)}_3 = \frac{1}{2} [\fsl{\epsilon},\fsl{\hat q}] ,\ \ 
     \Gamma^{(V)}_4 = \frac{1}{3!}[\fsl{\epsilon},\fsl{p},\fsl{\hat q}]
     \\  
& &  \Gamma^{(V)}_5 = (\epsilon \cdot p) ,\ \ 
     \Gamma^{(V)}_6 = \fsl{p} (\epsilon \cdot p) ,\ \ 
     \Gamma^{(V)}_7 = \fsl{\hat q}(p \cdot \hat q) 
           (\epsilon \cdot p) ,\ \ 
     \Gamma^{(V)}_8 = \frac{1}{2} [\fsl{p},\fsl{\hat q}]
           (\epsilon \cdot p) ,
\nonumber
\end{eqnarray}
where $[a,b,c] \equiv a[b,c] + b[c,a] + c[a,b]$.
For the axial-vector vertex we use 
\begin{eqnarray}
  \Gamma^{(A)}_1 &=& \fsl{\epsilon}\ \gamma_5 ,\ \ \ 
     \Gamma^{(A)}_2 = \frac{1}{2} [\fsl{\epsilon},\fsl{p}] 
                           \gamma_5  ,\ \ \ 
     \Gamma^{(A)}_3 = \frac{1}{2} [\fsl{\epsilon},\fsl{\hat q}]\ 
             (p \cdot \hat q)
                \ \gamma_5  ,\nonumber\\
     \Gamma^{(A)}_4 &=& \frac{1}{3!}[\fsl{\epsilon},\fsl{p},
             \fsl{\hat q}]
     \ \gamma_5 ,\ \ \ 
     \Gamma^{(A)}_5 = (\epsilon \cdot p)\ (p \cdot \hat q)\ 
       \gamma_5  ,\ \ \ 
     \Gamma^{(A)}_6 = \fsl{p} (\epsilon \cdot p)\ \gamma_5  ,
     \nonumber\\
     \Gamma^{(A)}_7 &=& \fsl{\hat q}\ (\epsilon \cdot p)\ 
            (p \cdot \hat q)
                \ \gamma_5  ,\ \ \ 
     \Gamma^{(A)}_8 = \frac{1}{2} [\fsl{p},\fsl{\hat q}]
            (\epsilon \cdot p)\
     (p \cdot \hat q)\ \gamma_5.
\end{eqnarray}
{}From the above choice of the bases, we can easily show that all the
invariant amplitudes $\chi_i^{(J)}$ are the even functions of 
$(p\cdot \hat{q})$ using the charge conjugation property of the
current.

In the present analysis we fix the frame of reference in such a way 
that only the zero component of the total momentum $q^\mu$ becomes 
non-zero.
Furthermore, we study the case where $q^\mu$ is in the space-like 
region.
Then, it is convenient to parameterize the total momentum $q^\mu$ as
\begin{equation}
  q^\mu = (i Q , 0 , 0 , 0 ) .
\end{equation}
For the relative momentum $p^\mu$, we perform 
the Wick rotation, and parameterize it by the real 
variables $u$ and $x$ as
\begin{equation}
  p \cdot q = - Q\  u \ ,\ \  p^2 = - u^2 - x^2 .
\end{equation}
Consequently, the invariant amplitudes $\chi^{(J)}_i$
become functions 
in $u$ and $x$:
\begin{equation}
  \chi_i^{(J)} = \chi_i^{(J)}(u,x;Q) .
\end{equation}
{}From the charge conjugation properties for the  
BS amplitude $\chi^{(J)}$ 
and the bispinor bases defined above, 
the invariant amplitudes $\chi_i^{(J)}(u,x)$ are shown to 
satisfy 
\begin{equation}
  \chi_i^{(J)}(u,x;Q) = \chi_i^{(J)}(-u,x;Q)\ .
\label{eq:even-property}
\end{equation}
Using this property of the invariant amplitudes, 
we rewrite Eq.~(\ref{eq:Pi_JJ}) as 
\begin{equation}
  \Pi_{VV}(Q^2) \ =\  \frac{N_c}{\pi^3} \int_0^\infty du 
  \int_0^\infty dx\ x^2 \left[ - \chi^{(V)}_1(u,x;Q) 
  + \frac{x^2}{3} \chi^{(V)}_6(u,x;Q) 
  \ \right],
\label{eq:PiVV}
\end{equation}
\begin{equation}
  \Pi_{AA}(Q^2) \ =\  \frac{N_c}{\pi^3} \int_0^\infty du 
  \int_0^\infty dx\ x^2 \left[\ \ \chi^{(A)}_1(u,x;Q) 
  - \frac{x^2}{3} \chi^{(A)}_6(u,x;Q) 
  \ \right].
\label{eq:PiAA}
\end{equation}
Here, we used the expanded form of the BS amplitude shown in 
Eq.~(\ref{eq:chi-expanded}) and carried out the three dimensional 
angle integration.

From Eqs.~(\ref{eq:PiVV}) and (\ref{eq:PiAA}), 
the quantity $\Pi_{VV} - \Pi_{AA}$ is expressed as
\begin{eqnarray}
  \Pi_{VV} - \Pi_{AA} &=&  
  \frac{1}{3} \sum_{\epsilon} \int \frac{d^4 p}{i (2
  \pi)^4} \frac{N_c}{2} \mbox{tr} \left[ 
  \fsl{\epsilon} \chi^{(J)}(p;q,\epsilon) - 
  \fsl{\epsilon} \gamma_5 \chi^{(A)}(p;q,\epsilon) 
  \right],\nonumber\\
  &=& 
  \frac{N_c}{\pi^3} \int_0^\infty du 
  \int_0^\infty dx\ x^2 \Bigg[ - \left( \chi^{(V)}_1(u,x;Q) 
  + \chi^{(A)}_1(u,x;Q) \right) \nonumber\\
  & &\ \ \ \ \ \ \ \ \ \ \ \ \ \ \ \ \ \ \ \ \ 
     + \frac{x^2}{3} \left( \chi^{(V)}_6(u,x;Q) 
  + \chi^{(A)}_6(u,x;Q) \right) 
  \ \Bigg].
\end{eqnarray}
We note that, although either $\Pi_{VV}$ or $\Pi_{AA}$ is
logarithmically divergent quantity, the difference
$\Pi_{VV} - \Pi_{AA}$ becomes 
finite due to the cancellation of the divergence ensured by
the chiral symmetry.

%%%%%%%%%%%%%%%%%%%%%%%%%%%%%%%%%%%%%%%%%%%%%%%%%%%%%%%%%%%%%%%%%%%%
%%%%%%%%%%%%%%%%%%%%%%%%%%%%%%%%%%%%%%%%%%%%%%%%%%%%%%%%%%%%%%%%%%%%
\section{Inhomogeneous Bethe-Salpeter equation}
\label{sec:IBS}

In this section we introduce the inhomogeneous Bethe-Salpeter
(IBS) equation 
from which we calculate the BS amplitude 
defined in the previous section.
We also show the numerical method for solving the IBS equation. 

%%%%%%%%%%%%%%%%%%%%%%%%%%%%%%%%%%%%%%%%%%%%%%%%%%%%%%%%%%%%%%%%%%%%
\subsection{IBS equation}

\begin{figure}
  \begin{center}
    \includegraphics[height=3cm]{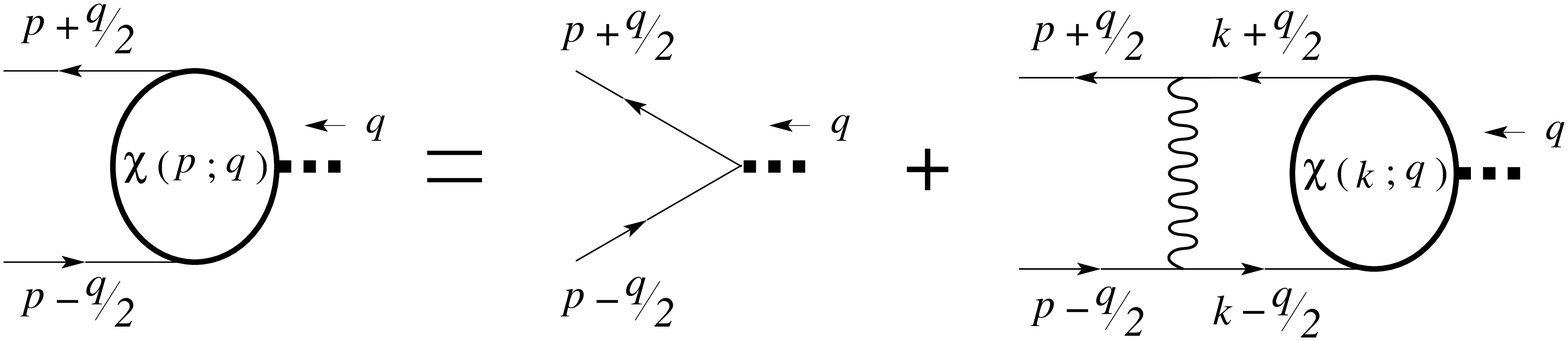}
  \end{center}
\caption{A graphical expression of the IBS equation in the 
(improved) ladder approximation.}
\label{fig:IBSeq}
\end{figure}

The IBS equation is the self-consistent equation for 
the BS amplitude $\chi^{(J)}$, and it is expressed as 
(see Fig.~\ref{fig:IBSeq} for graphical expression)
\begin{equation}
  T(p;q)\ \chi^{(J)}(p;q,\epsilon) \ \ =\ \ 
  \epsilon \cdot G^{(J)}\ 
  +\ K(p;k) \ast \chi^{(J)}(k;q,\epsilon) .
\label{eq:IBSeq}
\end{equation}
The kinetic part $T$ is given by 
\begin{equation}
  T(p;q) =   S_F^{-1}(p + q/2) \otimes  S_F^{-1}(p - q/2) \ ,
\label{def T}
\end{equation}
where $S_F$ is the full fermion propagator 
$i S_F^{-1}(p) = \fsl{p} - \Sigma(p) $. 
(Note that wave function renormalization factor $A(p)$ becomes unity 
when we use the Landau gauge.)
The BS kernel $K$ in the improved ladder approximation 
is expressed as
\begin{equation}
  K(p;k) \ =\    \frac{N_c^2 - 1}{2 N_c} \ 
         \frac{\bar{g}^2(p,k)}{- (p-k)^{2}}\ 
         \left( g_{\mu\nu} - \frac{(p-k)_\mu (p-k)_\nu}{(p-k)^2}
         \right) \cdot \gamma^\mu \otimes \gamma^\nu ,
\end{equation} 
where $\bar{g}(p,k)$ is the running coupling of QCD whose explicit 
form will be shown later.
In the above expressions we used
the tensor product notation
\begin{equation}
  (A \otimes B) \,\chi  =  A\, \chi\, B \ ,
\end{equation}
and the inner product notation 
\begin{equation}     K(p;k) \ast \chi^{(J)}(k;q,\epsilon) =   
  \int \frac{d^4 k}{i(2\pi)^4}\  K(p,k)\  \chi(k;q) \ .
\end{equation}

The mass function of the quark propagator is obtained from 
the SD equation:
\begin{equation}
  \Sigma(p) = K(p,k) \ast i S_F(p).
\end{equation}
It should be stressed that we must
use the same kernel $K(p,k)$ as that used in the 
IBS equation for consistency with the chiral 
symmetry~\cite{MN,Kugo:1992pr,Kugo:1992zg,Bando-Harada-Kugo}.

%%%%%%%%%%%%%%%%%%%%%%%%%%%%%%%%%%%%%%%%%%%%%%%%%%%%%%%%%%%%%%%%%%%%
\subsection{Numerical method for solving the IBS equation}

In this subsection we transform the IBS equation in 
Eq.~(\ref{eq:IBSeq}) into the form 
by which we can solve it numerically.

First, we introduce 
the conjugate bispinor bases defined by
\begin{equation}
  \bar\Gamma^{(J)}_i(p;q,\epsilon) \equiv
  \gamma_0 \Gamma^{(J)}_i(p^\ast;q,\epsilon)^\dag \gamma_0 \ .
\end{equation}
Multiplying 
these conjugate bispinor bases 
from left, taking the trace 
of spinor indices and summing over the polarizations, 
we rewrite
Eq.~(\ref{eq:IBSeq}) into the following form:
\begin{equation}
  T^{(J)}_{ij}(u,x) \chi^{(J)}_j(u,x) - \frac{1}{8 \pi^3} 
  \int_{-\infty}^{\infty}dv \int_0^\infty dy y^2 
  K^{(J)}_{ij}(u,x;v,y) \chi^{(J)}_j(v,y) 
  \ =\ I^{(J)}_i(u,x) ,
\end{equation}
where 
the summation over the index $j$ is understood, and
\begin{eqnarray}
  I_i^{(J)} &=& \sum_{\epsilon} 
        \mbox{tr} \left[ \bar\Gamma_i^{(J)}(p;q,\epsilon)
	\left( \epsilon \cdot G^{(J)}\right) \right] \ ,\\
  T^{(J)}_{ij}(u,x) &=& \sum_{\epsilon} 
        \mbox{tr} \left[ 
	\bar\Gamma_i^{(J)}(p;q,\epsilon) T(p;q) \Gamma_j^{(J)}
          (p;q,\epsilon)
	\right] ,\\
  K^{(J)}_{ij}(u,x;v,y) &=& \int_{-1}^{1} d\cos\theta 
         \ \sum_{\epsilon} 
        \mbox{tr} \left[ \bar\Gamma_i^{(J)}(p;q,\epsilon)
	K(p,k) \Gamma_j^{(J)}(k;q,\epsilon) \right] \ ,
\end{eqnarray}
with the real variables $v$ and $y$ introduced as
\begin{equation}
  k\cdot q = - v\ Q ,\ \ \  k\cdot p= - u v - x y \cos\theta .
\end{equation}
Here $\theta$ is the angle between the spatial components of 
$p_\mu$ and $k_\mu$.

Using the property of $\chi_i^{(J)}$ in Eq.~(\ref{eq:even-property}), 
we restrict the integration range as $v > 0$:
\begin{equation}
  \int dv K_{ij}(u,x;v,y) \chi_j^{(J)}(v,y) = 
	\int_{v > 0} dv \left[ K_{ij}(u,x;v,y) +
	 K_{ij}(u,x;-v,y) \right] \chi_j^{(J)}(v,y).
\end{equation}
Then, in the following, we treat
all the variables $u$, $x$, $v$ and $y$
as positive values.

To discretize the
variables $u$, $x$, $v$ and $y$
we introduce new variables 
$U$, $X$, $V$ and $Y$ as
\begin{eqnarray}
  u = e^{U}\ , & & x = e^{X} \ , \nonumber\\
  v = e^{V}\ , & & y = e^{Y} \ ,
\end{eqnarray}
and
set ultraviolet (UV) and infrared (IR) cutoffs as
\begin{equation}
  U, V \  \in \  [\lambda_U,\Lambda_U] ,\ \ 
  X, Y \  \in \  [\lambda_X,\Lambda_X] .
\end{equation}
We discretize the variables $U$ and $V$ 
into $N_{BS,U}$ points evenly, and $X$ and $Y$
into $N_{BS,X}$ points.
Then,
the original variables are labeled as
\begin{eqnarray}
  & & u_{[I_U]} = \exp\left[\lambda_U + D_U I_U \right], \ \ 
      x_{[I_X]} = \exp\left[\lambda_X + D_X I_X \right], \nonumber\\
  & & v_{[I_V]} = \exp\left[\lambda_U + D_U I_V \right], \ \ 
      y_{[I_Y]} = \exp\left[\lambda_X + D_X I_Y \right], \nonumber
\end{eqnarray}
where
$I_U, I_V = 0, 1, 2, \cdots (N_{BS,U}-1)$ and
$I_X, I_Y = 0, 1, 2, \cdots (N_{BS,X}-1)$.
The measures $D_U$ and $D_X$ are defined as
\begin{equation}
  D_U = \frac{\Lambda_U - \lambda_U}{N_{BS,U} - 1}\ ,\ \ 
  D_X = \frac{\Lambda_X - \lambda_X}{N_{BS,X} - 1} \ .
\end{equation}
As a result, the integration is converted into the 
summation:
\begin{equation}
  \int_{v > 0} y^2 \ dy\ dv\ \cdots  \ \ \  \Longrightarrow \ \ \  
  D_U D_V \sum_{I_V,I_Y} v y^3 \ \cdots.
\end{equation}
In
order to avoid integrable singularities
in the kernel $K(u,x;v,y)$ at $(u,x)=(v,y)$, we adopt 
the following four-splitting prescription~\cite{Aoki:1990yp}:
\begin{eqnarray}
  K_{ij}(u,x,v,y) \ \  &\Longrightarrow& \ \   \frac{1}{4} 
      \ [\ K_{ij}(u,x,v_+,y_+) + K_{ij}(u,x,v_+,y_-) \nonumber\\
  & & \ \ \ \ +\  K_{ij}(u,x,v_-,y_+) + K_{ij}(u,x,v_-,y_-)\ ] ,
\end{eqnarray}
where    
\begin{equation}
  v_\pm = \exp\left[V \pm \frac{D_U}{4}\right] , \ \ 
  y_\pm = \exp\left[Y \pm \frac{D_X}{4}\right] . 
\end{equation}
Now that all the variables have become discrete and 
the original integral equation (\ref{eq:IBSeq}) has
turned into a linear algebraic one, we are able to deal it 
numerically.

%%%%%%%%%%%%%%%%%%%%%%%%%%%%%%%%%%%%%%%%%%%%%%%%%%%%%%%%%%%%%%%%%%%%
%%%%%%%%%%%%%%%%%%%%%%%%%%%%%%%%%%%%%%%%%%%%%%%%%%%%%%%%%%%%%%%%%%%%
\section{Results and discussion}
\label{sec:results}

In this section, we show the results of calculations 
and give some discussions.
First, we discuss the dependence of the
results on the infrared structure 
of the running coupling.
Second, we compare the results with experimental values.

For solving the IBS and SD
equations, we have to fix the form of 
the running coupling $\bar{g}^2(p,k)$ which appears 
in the IBS and the SD equations.
We use the solution of the renormalization group equation 
for the QCD running coupling with one-loop approximation.
We regularize the infrared divergence of 
the one-loop running coupling by introducing the 
IR cutoff parameter $t_F\ ( >0 )$ as follows:
\begin{equation}
  \alpha(p,k) \equiv \frac{\bar{g}^2(p,k)}{4\pi} = 
    \alpha_0\ \frac{1}{
    \mbox{max}( t_F, t )},\ \ \ 
  t = \ln \left[(p_E^2+k_E^2)/\Lambda_{QCD}^2\right]
\label{eq:running}
\end{equation}
where $\alpha_0 = 12\pi/(11N_c - 2N_f)$ with 
$N_f$ being the number of flavors.
As for the argument of the running coupling $\bar{g}^2(p,k)$,
we used the angle averaged form $p_E^2 + k_E^2$ so that we can 
analytically
carry out the angle integration in the SD and the IBS equations.

\begin{figure}
  \begin{center}
    \includegraphics[height=7.5cm]{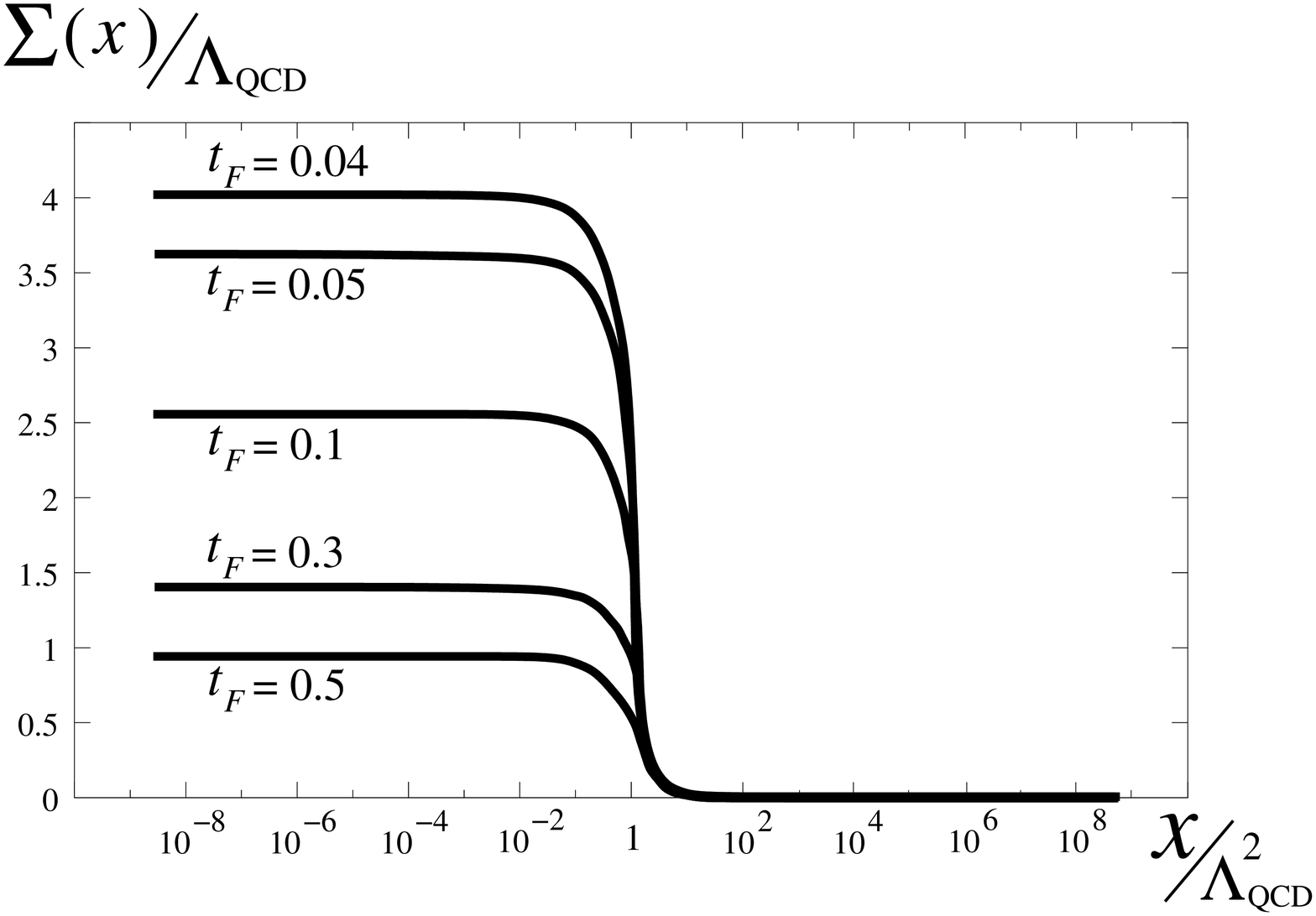}
  \end{center}
\caption{Mass functions obtained from the SD equation for several 
values of $t_F$.}
\label{fig:mass_function}
\end{figure}

In Fig.~\ref{fig:mass_function}, we plot the 
solutions of the SD equation
for several values of $t_F$.
For smaller value of $t_F$, the value of mass functions $\Sigma(x)$ 
in the infrared energy region becomes larger.
This is natural because smaller value of $t_F$ means 
larger running coupling in the infrared region.

When we solve the IBS equation, we use the following parameters:
\begin{equation}
 \left[ \lambda_U,\Lambda_U \right] = \left[ -5.5, 2.5 \right],
\end{equation}
\begin{equation}
 \left[ \lambda_X,\Lambda_X \right] = \left[ -2.5, 2.5 \right],
\end{equation}
\begin{equation}
  N_{BS,U} = N_{BS,X} = 30.
\end{equation}
These parameters are chosen so that the dominant supports 
always lie within the energy region between UV and IR cutoffs 
for all values of $t_F$ which we use in the present analysis
($t_F = 0.04$-$0.5$).

\subsection{$t_F$ dependence of $\Pi_{VV} - \Pi_{AA}$}

\begin{figure}
  \begin{center}
    \includegraphics[height=10cm]{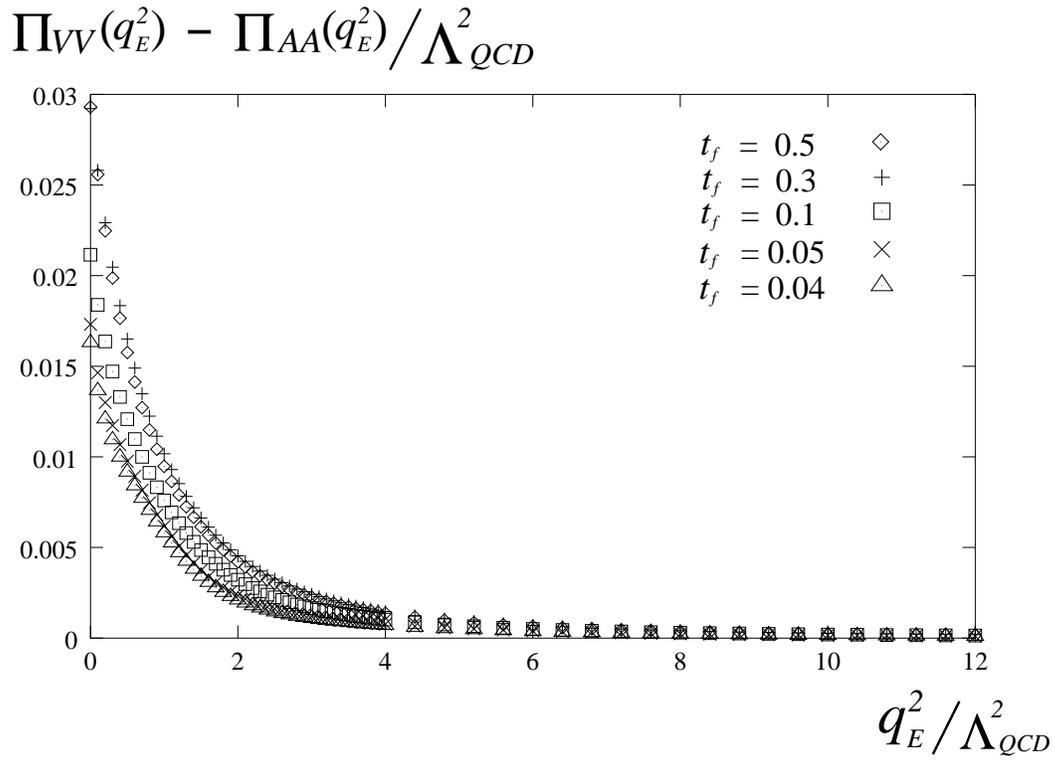}
  \end{center}
\caption{Resultant values of $\Pi_{VV}(Q^2) - \Pi_{AA}(Q^2)$ 
for several values of $t_F$.
Horizontal axis shows the square of the Euclidean momentum normalized
by the scale $\Lambda_{QCD}^2$.
$\Pi_{VV}(Q^2) - \Pi_{AA}(Q^2)$ in the vertical axis is also 
normalized by $\Lambda_{QCD}^2$}
\label{fig:Pi}
\end{figure}

In Fig.~\ref{fig:Pi}, 
we plot the resultant values of $\Pi_{VV}(Q^2) - \Pi_{AA}(Q^2)$ 
for several values of $t_F$.
Horizontal axis shows the square of the Euclidean momentum $Q$.
In this figure, $\Pi_{VV}(Q^2) - \Pi_{AA}(Q^2)$ and $Q^2$ 
are normalized by $\Lambda_{QCD}^2$, and plotted as the
dimensionless 
quantities.
As we mentioned before, 
either $\Pi_{VV}(Q^2)$ or $\Pi_{AA}(Q^2)$
is logarithmically divergent.
These divergences are expected to be canceled in  $\Pi_{VV}(Q^2) -
\Pi_{AA}(Q^2)$ ensured by the chiral symmetry.
Figure~\ref{fig:Pi} shows this cancellation actually 
occurs in the present calculation.~\footnote{
 If the quadratic divergences were not canceled, 
 $\Pi_{VV}-\Pi_{AA}$ would become 
 $\frac{\Lambda_{QCD}^2}{8\pi^2}\log\Lambda^2 \sim 0.1
 \Lambda_{QCD}^2$.
 Figure~\ref{fig:Pi} shows that $\Pi_{VV}-\Pi_{AA}$ is about 
 $0.03\Lambda_{QCD}^2$ at biggest, so that the divergences are
 actually canceled with each other.
}
{}From the results shown in Fig.~\ref{fig:Pi}, 
we can see that $\Pi_{VV}(Q^2) - \Pi_{AA}(Q^2)$ 
is dependent on the value of $t_F$.
In other words, 
$\Pi_{VV}(Q^2) - \Pi_{AA}(Q^2)$ is 
sensitive to the infrared structure of the QCD running coupling.

Once we obtain $\Pi_{VV}(Q^2) - \Pi_{AA}(Q^2)$, 
we can calculate the QCD $S$ parameter, $f_\pi$ and $\Delta m_\pi^2$ 
by using Eqs.~(\ref{eq:S_parameter}), (\ref{eq:fpi}) and 
(\ref{eq:mass_diff}).
In table \ref{tab:dimensionless}, we list the resultant values 
of these quantities for several values of $t_F$.
Here, we normalized dimensionfull quantities by $\Lambda_{QCD}$, 
and expressed them as dimensionless quantities.
\begin{table}
\begin{center}
 \begin{tabular}{|c||c|c|c|c|c|c|}  \hline
$t_F$ & 0.04
 &0.05 & 0.1 
 & 0.2 & 0.3 
 & 0.5 \\ \hline\hline 
   S 
       & 0.33  & 0.33  & 0.35 & 0.39 & 0.43 & 0.47 \\ \hline 
   $f_\pi$ $/\Lambda_{QCD}$            
       & 0.128  & 0.132  & 0.145 & 0.162 & 0.171 & 0.171 \\ \hline 
   $\Delta m_\pi^2$ $/\Lambda_{QCD}^2$   
       & 0.00201  & 0.00203  & 0.00212  & 0.00215  & 0.00212  
       & 0.00201 \\ \hline 
 \end{tabular}
\end{center}
\caption{Resultant values of the QCD $S$ parameter, $f_\pi$ and $\Delta
m_\pi^2$ for several values of $t_F$.
Dimensionfull quantities are normalized by $\Lambda_{QCD}$ and
expressed as dimensionless quantities.}
\label{tab:dimensionless}
\end{table}
These results show that the {\it QCD $S$ parameter and 
$f_\pi/\Lambda_{QCD}$ are 
sensitive to the value of IR cutoff parameter $t_F$ while 
$\Delta m_\pi^2/\Lambda_{QCD}^2$ is less sensitive} to it.
This is natural because 
the QCD $S$ parameter and $f_\pi$ are directly related to the 
infrared quantities as can be easily seen from 
Eqs.~(\ref{eq:S_parameter}) and (\ref{eq:fpi}),
while the infrared dependence of the
integration of  $\Pi_{VV}(Q^2) - \Pi_{AA}(Q^2)$ 
in Eq.~(\ref{eq:mass_diff}) is compensated by 
$f_\pi^2 (= \Pi_{VV}(0) - \Pi_{AA}(0))$ in the denominator.

Table~\ref{tab:dimensionless} 
shows that, when we take the value of $t_F$ smaller, 
the value of $f_\pi/\Lambda_{QCD}$ becomes smaller.
Here, smaller value of $t_F$ means larger value of infrared running
coupling, or, in other words, strong infrared dynamics.
At first sight, this behavior seems strange because $f_\pi$ is 
the order parameter of the chiral symmetry breaking, and 
it is expected to reflect the magnitude of the chiral symmetry
breaking.
One might think that $f_\pi/\Lambda_{QCD}$ becomes larger 
for smaller value of $t_F$.
However, it is not the case.
In the present analysis, 
$f_\pi$ is not necessarily proportional to the magnitude of the 
chiral symmetry breaking because QCD with $N_f=3$ in the vacuum
is very far from 
the chiral phase transition point.
To understand this behavior, let us see the Pagels-Stokar (PS) 
formula~\cite{Pagels:hd}
\footnote{
 We checked that the difference between $f_\pi$ calculated
 from the IBS equation and the PS formula is less than 10 \%
 for all values of $t_F$.
 We also checked that qualitative feature of $t_F$ dependence of
 $f_\pi$
 calculated from the PS formula is not different from $f_\pi$ obtained 
 from the IBS equation.
}:
\begin{equation}
  f_\pi^2 \ = \ \frac{N_c}{4\pi^2} \int_0^\infty dx\  
       x\ \frac{\ \Sigma^2(x) \ -\  \frac{ x }{4} \ \frac{d}{dx}
       \left[\  \Sigma^2(x) \ \right]\ }
       {\left[\  x \ +\  \Sigma^2(x) \ \right]^2} .
\label{eq:PS}
\end{equation} 
As shown in Fig.~\ref{fig:mass_function}, the mass functions 
$\Sigma(x)$ are almost constant for $0 < x=Q^2 < \Lambda_{QCD}^2$ 
and suddenly drops at $x = \Lambda_{QCD}^2$ for all values of $t_F$.
So we can cut the integration in Eq.~(\ref{eq:PS}) at 
$x = \Lambda_{QCD}^2$ and drop the derivative of mass
function.
We also drop $x$ in the denominator since $x$ satisfies 
$x \ll \Sigma(x)$ for the relevant integral region
($0 < x < \Lambda_{QCD}^2$).
Then we can approximate Eq.~(\ref{eq:PS}) as 
\begin{equation}
  f_\pi^2 \ \simeq \ \frac{N_c}{4\pi^2} \int_0^{\Lambda_{QCD}^2} dx\  
       x\ \frac{ 1 }
       { \Sigma^2(0) } 
  \ \propto\ \frac{\Lambda_{QCD}^4}{\Sigma^2(0)}.
\label{eq:PS-app}
\end{equation} 
{}From this, we can easily understand that, when 
infrared dynamics becomes strong, i.e., $\Sigma(0)$ becomes large, 
$f_\pi$ becomes small.

For comparison, we consider the decay constant near 
the chiral phase transition point.
In this case, $\Sigma(x)$ becomes very small and Eq.~(\ref{eq:PS}) 
is well approximated by 
\begin{equation}
  f_\pi^2 \ = \ \frac{N_c}{4\pi^2} \int_{\Sigma^2(0)}^\infty   
       \frac{dx}{x}\ \Sigma^2(x).
\label{eq:PS-crit}
\end{equation} 
From this, we can see that 
if the running coupling, and then the mass function 
becomes smaller,
$f_\pi$ also becomes smaller.
Thus near the critical point, order parameter $f_\pi$ is actually 
proportional to the magnitude of the chiral symmetry breaking.
We can actually see this behavior in the
large $N_f$ QCD near the critical
point (see Figs.~4, 5 in Ref.~\cite{Harada:2003dc}).

\subsection{Comparison with experiments}

In this subsection, we compare our results with experiments.
So far in this paper, all the dimensionfull quantities 
were normalized by $\Lambda_{QCD}$ and dealt as dimensionless 
quantities.
Here, we introduce the ``physical energy scale'' 
by setting $\Lambda_{QCD}$ in a way that it reproduces 
the experimental value of $f_\pi$. Since $f_\pi$ is sensitive to the
infrared 
parameter, fixing $f_\pi$ implies that  $\Lambda_{QCD}$ in turn
appears sensitive 
to the infrared parameter and so does the $\Delta m_\pi^2$ which was
shown  
insensitive to the infrared parameter as far as $\Lambda_{QCD}$ is
fixed.  
We use $f_\pi = 92.4~{\rm MeV}$ as an input to fix $\Lambda_{QCD}$
in the present analysis.\footnote{
 In the chiral symmetry limit of three flavors $m_u=m_d=m_s=0$, 
 the pion decay constant is estimated as $f_\pi
 =86.4\pm 0.26 \, {\rm MeV}$~\cite{Harada:2003jx}. 
 In this paper we use instead the
 physical value only for fixing $\Lambda_{QCD}$  as a reference scale,
 which is up to the ``scale ambiguity'' in  the SD and BS approach.
}

Table~\ref{tab:results-1} shows comparison of the resultant 
values of $\Delta m_\pi^2$ and $S$ 
for several values of $t_F$ with experimental values.
\begin{table}
\begin{center}
 \begin{tabular}{|c||c|c|c|c|c|c||c|}  \hline
   $t_F$ & 0.04 & 0.05 & 0.1 & 0.2 & 0.3 & 0.5 & Exp.\\ \hline\hline 
   $\Delta m_\pi^2$ $({\rm MeV^2})$   
       & 1050  & 1003  & 855  & 698  & 620  & 585 & 1261.1 \\ \hline 
   $S$ 
       & 0.33  & 0.33  & 0.35 & 0.39 & 0.43 & 0.47 & 0.32 $\pm$ 0.04
   \\ \hline 
   $f_\pi$ $({\rm MeV})$\ (input) 
       & 92.4  & 92.4  & 92.4 & 92.4 & 92.4 & 92.4 & 92.4 \\ \hline 
 \end{tabular}
\end{center}
\caption{Resultant values of
$\Delta m_\pi^2$, $S$ and $f_\pi$ for several values of
$t_F$.
Here, we used $f_\pi = 92.4~{\rm MeV}$ as an input to introduce 
the physical energy scale.
Experimental values of $\Delta m_\pi^2$ and $f_\pi$ are given in  
Ref.~\cite{Hagiwara:fs}, and $S$ is 
obtained from the value of the $L_{10}$ given in 
Ref.~\cite{Harada:2003jx} through
$S = -16\pi\left[ L_{10}(\mu) + \frac{1}{192\pi^2}
\left(\ln\frac{m_\pi^2}{\mu^2} + 1 
\right)
\right]$.}
\label{tab:results-1}
\end{table}
$\Delta m_\pi^2$, $S$ and $f_\pi$ are directly calculated from 
$\Pi_{VV}(Q^2) - \Pi_{AA}(Q^2)$ by using 
Eqs.~(\ref{eq:mass_diff}), (\ref{eq:S_parameter}) 
and (\ref{eq:fpi}).
From table~\ref{tab:results-1}, we can see that if we change the 
value of $t_F$, values of $\Delta m_\pi^2$ and $S$ move to
opposite directions:
When we decrease the value of $t_F$, 
$\Delta m_\pi^2$ becomes large, while 
$S$ becomes small.
For the choice $t_F = 0.5$, 
which was 
adopted in Ref.~\cite{Kugo:review,Aoki:1990eq,Aoki:1990yp} 
as a good regularization in the calculations of the
masses and decay constants of $\rho$ and $a_1$ mesons, 
we obtained
$\Delta m_\pi^2 = 585\, {\rm MeV^2}$  
and $S = 0.47$.
Both of these values are far from their experimental values 
$\Delta m_\pi^2 = 1261.1\, {\rm MeV^2}$ and $S = 0.29 - 0.36$.

However, as we decrease the value of $t_F$, 
both $\Delta m_\pi^2$ and $S$ move toward
their experimental values.
At $t_F = 0.04$, they become
\begin{eqnarray}
 \Delta m_\pi^2 &=& 1050\, {\rm MeV^2}\,
 ,\\ 
 S &=& 0.33\, , 
\end{eqnarray}
in good agreement with the experimental value of $S$ and 
in rough agreement with that of $\Delta m_\pi^2$.
In the case of $t_F = 0.04$  we take 
\begin{equation}
\Lambda_{QCD} = 724 \, {\rm MeV}
\end{equation} 
in order to set $f_\pi = 92.4 \,{\rm MeV}$.
This value is much larger than the conventional value of
$\Lambda_{QCD}$ in the
$\overline{\rm MS}$ scheme~\cite{Buras}:
\begin{equation}
\Lambda_{\rm QCD}^{(3)} = 300-450\,{\rm MeV} \,.
\end{equation}
Such a problem already occurred in the previous 
works~\cite{Kugo:review,Aoki:1990eq,Aoki:1990yp},  
\begin{equation}
\Lambda_{QCD} \simeq 500\, {\rm MeV} \, ,
\end{equation}
which 
roughly corresponds to $t_F = 0.5$ in the present analysis.
This reflects the ``scale ambiguity''~\cite{BLM},  
as was recently pointed out in Ref.~\cite{Hashimoto:2002px}
where an interesting method was proposed  to solve it by using 
the effective coupling.

{}From these results, we conclude that, 
when we calculate $f_\pi$, $\Delta m_\pi^2$ and $S$ 
by using the BS equation with the improved ladder approximation, 
the infrared cutoff parameter $t_F$ should be taken 
smaller than values used so far in the previous 
works~\cite{Kugo:review,Aoki:1990eq,Aoki:1990yp}.
This means that the running coupling in the infrared 
energy region should be taken larger in order to 
reproduce experimental values of $f_\pi$, $\Delta m_\pi^2$ and $S$ 
at the same time in the calculations 
based on the BS equation with the improved ladder approximation.
(When we take $t_F = 0.04$, the value of the running coupling in the
infrared energy region becomes about $\alpha \sim 35$.)

Now let us look at what other quantities can be obtained 
from $\Pi_{VV} - \Pi_{AA}$.
As for $\rho$ meson and $a_1$ meson, 
we derive their masses and decay constants by fitting them
to the calculated $\Pi_{VV}(Q^2) - \Pi_{AA}(Q^2)$ 
using the pole saturated form 
of $\Pi_{VV}(Q^2) - \Pi_{AA}(Q^2)$.  
Here, we use the following simplest version of the pole saturated 
form as a fitting function:
\begin{equation}
  \left[ \Pi_{VV}(Q^2) - \Pi_{AA}(Q^2) \right]_{\mbox{fit}}\ = \ 
    - \frac{Q^2 f_\rho^2}{M_\rho^2 + Q^2} + f_\pi^2 
    + \frac{Q^2 f_{a_1}^2}{M_{a_1}^2 + Q^2}.
\end{equation}
\begin{table}
\begin{center}
 \begin{tabular}{|c||c|c|c|c|c|c||c|}  \hline
   $t_F$ & 0.04 & 0.05 & 0.1 & 0.2 & 0.3 & 0.5 & Exp.\\ \hline\hline 
   $M_\rho$ $({\rm MeV})$           
       & 730 & 710 & 716 & 645 & 613 & 612 & 770 - 772 \\ \hline 
   $f_\rho$ $({\rm MeV})$           
       & 150 & 146 & 151 & 150 & 150 & 151 & 154 \\ \hline 
   $M_{a_1}$ $({\rm MeV})$           
       & 1186 & 1146 & 1012 & 908 & 861 & 859 & 1190 - 1270 \\ \hline 
   $f_{a_1}$ $({\rm MeV})$           
       & 115 & 109 & 119 & 118 & 119 & 123 & 144 \\ \hline 
 \end{tabular}
\end{center}
\caption{Resultant best fitted values of 
masses and decay constants of $\rho$ and $a_1$ mesons 
for several values of $t_F$.
Experimental values for each quantities are also listed.
Experimental values of $M_\rho$, $f_\rho$ and $M_{a_1}$ are given in 
Ref.~\cite{Hagiwara:fs}, and $f_{a_1}$ is given in 
Ref.~\cite{Isgur:vm}.
}
\label{tab:results-2}
\end{table}
Resultant best fitted values are listed in 
table~\ref{tab:results-2} together with the
experimental values.
From the results shown in this table, 
we can see that the
masses and the decay constants of $\rho$ and 
$a_1$ mesons are not so much dependent on the 
infrared cutoff parameter 
$t_F$ compared with $t_F$ dependence of 
$\Delta m_\pi^2$ and $S$.
When we choose $t_F = 0.04$, for which the values of
$\Delta m_\pi^2$ and $S$ are predicted in good agreement with
experiments, we obtained 
\begin{eqnarray}
 M_\rho &=& 730~{\rm MeV}\, ,\\
 f_\rho &=& 150~{\rm MeV}\, ,\\
 M_{a_1}&=& 1186~{\rm MeV}\, ,\\
 f_{a_1}&=& 115~{\rm MeV}\, ..
\end{eqnarray}
These are also in good agreement with experiments.
This confirms the reliability of calculations 
in the present analysis 
based on the improved ladder BS equation with small 
infrared cutoff parameter ($t_F \sim 0.04$).

%%%%%%%%%%%%%%%%%%%%%%%%%%%%%%%%%%%%%%%%%%%%%%%%%%%%%%%%%%%%%%%%%%%%
%%%%%%%%%%%%%%%%%%%%%%%%%%%%%%%%%%%%%%%%%%%%%%%%%%%%%%%%%%%%%%%%%%%%
\section{Summary and Discussions}
\label{sec:summary}
We have  calculated
the  $\pi^+-\pi^0$ mass difference $\Delta m_\pi^2$
on the same footing as the pion decay constant
$f_\pi$ and the QCD $S$ parameter through 
the vector and the axial-vector current 
correlators. For the calculations we
solved inhomogeneous BS
equations for the vector and axial-vector vertex functions
together with SD equation for the quark mass function within
 the improved ladder approximation in the Landau gauge.
We also obtained the masses and the decay constants of $\rho$ 
and $a_1$ mesons by fitting them to 
$\Pi_{VV}(Q^2) - \Pi_{AA}(Q^2)$ using the pole saturated form.

We showed that all of these quantities
are simultaneously fit in agreement with the experiments 
when we take the IR cutoff parameter $t_F=0.04$, 
which corresponds to taking 
\begin{equation}
\Lambda_{\rm QCD} \simeq 724 \, {\rm MeV}\,.
\end{equation}
This parameter choice of $t_F$ is fairly smaller than the ones 
investigated in the previous
works~\cite{Aoki:1990eq,Aoki:1990yp,Roberts:2000aa,Harada:1994ni}, 
which corresponds to 
\begin{equation}
\Lambda_{\rm QCD} \simeq 500 \, {\rm MeV}\, .
\end{equation}
Either value of $\Lambda_{\rm QCD}$ above is substantially larger than
that in 
the $\overline{\rm MS}$ scheme~\cite{Buras}:
\begin{equation}
\Lambda_{\rm QCD}^{(3)} = 300-450\,{\rm MeV} \,.
\end{equation}
Our results imply 
that the running coupling of QCD 
in the IR region should be taken larger  
than that considered so far, 
when we calculate not only $f_\pi$ but also the QCD $S$ parameter 
and $\Delta m_\pi^2$ 
in the framework of
the BS and the SD equations in the improved 
ladder approximation.
Apparently, choosing the IR cutoff parameter
corresponds
to exploiting the ``scale ambiguity''~\cite{BLM} 
of $\Lambda_{\rm QCD}$ in
the SD and BS equations as was recently 
emphasized in Ref.~\cite{Hashimoto:2002px} which proposed an
interesting method to resolve the scale ambiguity in the 
the calculation of $f_\pi$.

In order to establish the approach of the SD and the BS equations
in the improved ladder approximation, it is certainly desirable to
apply the method of Ref.~\cite{Hashimoto:2002px} 
to our case, namely the calculations of  $\Delta m_\pi^2$ 
and QCD $S$ parameter
as well as $f_\pi$. This will be done in future work. 

The success of the analysis based on the BS and the SD equations 
 in the real-life QCD in this paper
well motivate us to apply this method for studies of 
other strong coupling gauge theories.
One of the most interesting examples of such 
is the large $N_f$ QCD
(see, e.g., Refs.~\cite{Banks:nn,Appelquist:1996dq,
Miransky:1996pd,lattice,OZ,VS,Harada:2003dc}), 
which is a QCD with large number of massless flavors (but not too
large 
number as to destroy the asymptotic freedom).
It is expected that the
chiral symmetry gets restored for a certain large number of massless
 flavors, 
which was in fact confirmed by the lattice simulations~\cite{lattice}.

It was argued~\cite{Harada:2000kb,Harada:2003jx} based on the
effective field 
theory of HLS~\cite{BKUYY,BKY:NPB,BKY:PTP} 
that this chiral restoration of the large $N_f$ QCD
is accompanied by the massless vector meson degenerate with the NG
boson  
(pseudoscalar meson) as the chiral partners
(``Vector Manifestation'' of chiral symmetry). It
was further argued ~\cite{Harada:2003xa} that the HLS model even in
the real-life QCD behaves as a little Higgs  
model~\cite{Arkani-Hamed:2001nc} with two sites and two links, whose 
locality of theory space forbids the quadratic divergence in
$\Delta m_\pi^2$ which is nothing but the $({\rm mass})^2$ of the
Higgs   
in the framework of the Little Higgs 
models. In the large $N_f$ QCD
$\Delta m_\pi^2$ becomes small compared with $f_\pi^2$ near the 
chiral restoration point, which  may suggest that 
the large $N_f$ QCD may be considered as
a UV completion 
of the Little Higgs model.

On the other hand, we found in the previous
paper~\cite{Harada:2003dc}, by the 
explicit calculation based on the homogeneous BS 
and the SD equations in the improved
ladder approximation, 
that masses of the  
scalar, vector, and axial-vector mesons in the large $N_f$ QCD
are proportional to $f_\pi$ and
vanish at the chiral restoration point, which implies somewhat
different 
manifestation than the Vector Manifestation in the HLS model.
(This behavior was also conjectured in 
Ref.~\cite{Chivukula:1996kg}.)

Thus it would be very useful to clarify the situation to calculate 
the $S$ parameter and $\Delta m_\pi^2$ in large $N_f$ QCD which
have not yet been  calculated directly from QCD. In the forthcoming
paper we shall calculate $\Delta m_\pi^2$ and $S$ parameter
in large $N_f$ QCD by the same method presented in this paper.

%%%%%%%%%%%%%%%%%%%%%%%%%%%%%%%%%%%%%%%%%%%%%%%%%%%%%%%%%%%%%%%%%%%%
%%%%%%%%%%%%%%%%%%%%%%%%%%%%%%%%%%%%%%%%%%%%%%%%%%%%%%%%%%%%%%%%%%%%

\section*{Acknowledgments}
We would like to thank Yoshio Kikukawa and Masaharu Tanabashi
for discussions. M.K. would like to thank Thomas Appelquist for 
warm hospitality and 
valuable discussion during his stay at Yale University.
This work was supported in part by 
the JSPS Grant-in-Aid for the Scientific Research 
(B)(2) 14340072 (K.Y. and M.H.) and by 
the 21st Century COE Program 
of Nagoya University provided by JSPS (15COEG01).

%%%%%%%%%%%%%%%%%%%%%%%%%%%%%%%%%%%%%%%%%%%%%%%%%%%%%%%%%%%%%%%%%%%%
%%%%%%%%%%%%%%%%%%%%%%%%%%%%%%%%%%%%%%%%%%%%%%%%%%%%%%%%%%%%%%%%%%%%

%%%%%%%%%%%%%%%%%%%%%%%%%%%%%%%%%%%%%%%%%%%%%%%%%%%%%%%%%%%%%%%%%%%%
%%%%%%%%%%%%%%%%%%%%%%%%%%%%%%%%%%%%%%%%%%%%%%%%%%%%%%%%%%%%%%%%%%%%

\end{document}